\documentstyle[preprint,revtex]{aps}
\begin{document}
\draft
\begin{title}
Magnetic Properties of Undoped $C_{60}$
\end{title}
\author{D.~Coffey$^a$ and S.~A.~Trugman$^b$}
\begin{instit}
$^a$ Center for Materials Science, MS-K765, Los Alamos National Laboratory,
Los Alamos, NM 87545
\end{instit}
\begin{instit}
$^b$ Theoretical Division, MS-B262, Los Alamos National Laboratory,
Los Alamos, NM 87545
\end{instit}
\begin{abstract}
The Heisenberg antiferromagnet,
which arises from the large $U$ Hubbard model, is investigated
on the $C_{60}$ molecule and other fullerenes.
The connectivity of $C_{60}$ leads to an exotic
classical ground state with nontrivial topology.
We argue that there is no phase transition in the Hubbard model
as a function of $U/t$, and thus the large $U$ solution
is relevant for the physical case of intermediate coupling.
The system undergoes a first order metamagnetic
phase transition.
We also consider the S=1/2 case using perturbation theory.
Experimental tests are suggested.
\end{abstract}
\narrowtext

The $C_{60}$ molecule (buckminsterfullerene) has carbon
atoms arranged like the vertices of a soccer ball.\cite{kroto}
We consider a neutral
molecule.
The active orbitals are one radial p-orbital for each carbon atom.
When the long range Coulomb or on-site
Hubbard repulsion is large compared with the
nearest neighbor hopping $t$, the molecule has essentially one electron
in each orbital.  There is an antiferromagnetic Heisenberg
spin-spin interaction between nearest neighbors, caused
by superexchange.  The exchange constant
$J = t^2 / \Delta$, where $\Delta$ is the energy difference
between a state with one electron in each orbital, and a state
with two electrons in one orbital and none in a neighboring orbital.
Including on-site $U$ and nearest neighbor $V$ interactions,
$\Delta = U-V$.
The derivation is similar to
that of the $t-J$ model used in connection with
the superconducting cuprates.

Estimates for
real $C_{60}$ are that $U$ is approximately 9 $eV$ and
$t$ is 2 to 3 $eV$.\cite{kivelson,stollhoff}  The real molecule
is thus in the intermediate $U$ regime, and not in the large $U$ limit
for which we calculate.
The $C_{60}$ molecule is too large to numerically solve for
the intermediate $U$ ground state.
There is evidence, however, that there is no phase transition for
the Hubbard model in the $C_{60}$ geometry as a function of
$U/t$.
The spin correlations
for intermediate $U/t$ are then expected to be qualitatively similar,
but smaller in magnitude, to those for large
$U/t$.
The evidence for a lack of a phase transition is as follows:
For a finite quantum system at zero temperature, a phase
transition as a function of the parameters occurs only if
the quantum numbers of the ground state change.
The only quantum numbers for this problem are spin $S$
and angular momentum $L$
(technically, what remains of $L$ under the symmetry
group of the icosahedron).  $S=0$ and $L=0$ in the limit
$U/t = 0$, and
probably also in the limit $U/t \rightarrow \infty$.\cite{largeu}
The simplest (and we believe correct) hypothesis is that
there is no phase transition as a function of $U/t$.
It may be useful in this regard
to consider the simple example of the two-site
Hubbard model with two electrons.  Mean field theory gives a
phase transition with sublattice magnetization developing at finite
$U/t$, which suggests that the large $U$ limit is not continuously
connected to the small $U$ limit.  The trivial exact solution, however,
makes it clear that there is in fact no phase transition, that local
moments develop continuously, and that the
spin correlations of the large $U$ limit develop continuously as
$U$ increases.


All carbon atoms on the $C_{60}$ molecule are equivalent, but there
are two slightly different bond lengths,
$1.45 \AA$ for the pentagon bonds and $1.40 \AA$ for the non-pentagon
bonds.\cite{yannoni}
The magnetic exchange constant for two neighboring sites
on the same pentagon is $J_1$.  The constant connecting a site
on one pentagon with a nearest neighbor on another pentagon is $J_2$.
The Hamiltonian is

\begin{equation}
H  = {J_1} \sum _ {<j,k>} ^ p \vec \sigma _j \cdot \vec \sigma _k
+ {J_2} \sum _ {<j,k>} ^ {np} \vec \sigma _j \cdot \vec \sigma _k ,
\label{ham}
\end{equation}
where the first sum is over the 60 pentagon bonds,
and the second over the 30 non-pentagon bonds.
$J_2$ is expected to be slightly larger than $J_1$, because the non-pentagon
bonds are shorter.
We first treat the Hamiltonian classically, so that $\vec \sigma$
is a classical unit vector.
This is the $S \rightarrow \infty$
limit.  (We consider the quantum $S = 1/2$ case at the end of the paper.)
The pentagons are frustrated, and cannot achieve a classical energy
of $-J_1$ per bond.  The ground state of an isolated pentagon
(5 spin system)
has all spins coplanar and an energy of $J_1 \cos ( 4 \pi / 5 ) = -.80902 J_1$
per bond.
One immediately obtains a {\it lower} bound for the energy of the entire ball,
which is $E_b = 60 J_1 \cos ( 4 \pi / 5 ) - 30 J_2$,
where we have used the fact that non-pentagon bonds cannot have an energy
lower than $-J_2$.

It would appear that the ground state energy of Eq. (\ref{ham}) cannot
achieve the lower bound $E_b$, because in simple trial states,
connected pentagons interfere with each other and increase the energy.
The classical ground state configuration was found numerically by minimizing
the energy over the spin variables $\{ \theta _ i , \phi _ i \},~i=1,60.$
See Fig. (\ref{spin}).
Surprisingly, the ground state energy is equal to the lower bound $E_b$.
The spin configuration is, however, nontrivial.
The 5 spins in any given pentagon are coplanar, but the
spins in a neighboring pentagon lie in a different plane.
(A bond connects neighboring pentagons, but the spins on
either end of the bond are precisely antiparallel, so that
the spin planes need not be identical.)
The ground state configuration, which has zero net moment,
is the same for all positive $J_1$ and $J_2$.
In addition to the obvious global rotational symmetry of the ground state,
there is a discrete parity symmetry whereby each spin
$\vec \sigma _i  \rightarrow -\vec \sigma _i$.

Since the
spins in any given pentagon are coplanar, $\hat n$
for a pentagon can be defined
to be the normal to the spin plane.
Define a second vector $\hat m$ for the pentagon,
which is normal to the physical plane of the pentagon
(a unit vector pointing
away from the center of the ball).
For a particular global spin rotation, the set of normals to
the spin planes for the 12 pentagons $\{\hat n_j\}$ is a nontrivial permutation
of the set of physical normals $\{\hat m_j\}$.
Let the global spin rotation be such that for the pentagon on the north pole,
$\hat n_1$ = $\hat m_1$.
The pentagon on the north pole is surrounded by
a first ring of 5 nearest neighbor pentagons, a second ring of 5 second nearest
neighbors, and one pentagon on
the south pole.
Any of the 5 pentagons in the first ring have an $\hat n$
equal to one of the $\hat m$ for a pentagon in the second ring.
Using the pentagon numbering convention of Fig. (\ref{spin}),
$\hat n _ i  = \hat m _ {j(i)}$, with
$(~j(1), j(2), \dots, j(12) ~) = $
(1, 7, 9, 11, 8, 10, 6, 3, 5, 2, 4, 12).
The solid angle subtended by the $\{\hat n_j\}$ for 3
pentagons that are mutual nearest neighbors
is 7 times larger than the solid
angle subtended by their $\{\hat m_j\}$.
A topological skyrmion number is the number of times one sphere covers another
sphere, much as a vortex number is the number of times one circle
covers another circle.
The skyrmion number for the field $\hat n$ is 7,
for both of the parity related ground states.\cite{skyrmion}
Given that the Hamiltonian is
so simple, this exotic ground state arises because of the connectivity of
the $C_{60}$.\cite{coxeter}


The classical antiferromagnet on
$C_{12}$, $C_{20}$, $C_{70}$, and $C_{84}$
has also been solved numerically.
The truncated tetrahedron $C_{12}$ is a smaller system with properties similar
to $C_{60}$.\cite{white}
It has 4 triangles and 4 hexagons.
The classical lower bound for the truncated tetrahedron,
$E_b^{(12)} = 12 J_1 \cos ( 2 \pi / 3 ) - 6 J_2$ is achieved by
the ground state, which
has a skyrmion number 1.

A large number of $C_n$ fullerenes
have been isolated.\cite{curl}
These compounds are closed (have the topology of a sphere), with
12 pentagons and a variable number of hexagons $n_h = (n/2)-10$.
The average frustration decreases as the number of
hexagons increases.
We have calculated the classical ground state for
the smallest molecule in this series, the dodecahedron $C_{20}$,
for the most
stable $D_{5h}$ isomer of $C_{70}$,
and for the $T_d$ isomer of
$C_{84}$.\cite{raghavachari}
The $C_{20}$ molecule has not
been synthesized, while
$C_{70}$ and $C_{84}$ are produced
in carbon arcs.

The ground state energy for the dodecahedron
is $-22.360680 J$, which does not reach the bound
$E_b^{(20)} = 30 J \cos ( 4 \pi / 5 )= -24.270510 J$.
The pentagons interfere with each other, and prevent the system
from reaching $E_b^{(20)}$.  The skyrmion number of the ground
state is 7.
Assuming that all $C_{70}$ bonds
have the same coupling $J$, the ground
state energy of $C_{70}$ is $-93.346473 J$, which is slightly
higher than the lower bound
$E_b^{(70)} = [ 60  \cos ( 4 \pi / 5 ) -45 ] J= -93.541020 J.$
The skyrmion number of the $C_{70}$ ground state is undefined.
There are pentagons neighboring across the equator whose spin plane normals
are precisely antiparallel, resulting in an undefined solid angle.
Many (but not all) states that differ infinitesimally from
the ground state have skyrmion number 7.
The lowest-lying metastable configuration for $C_{70}$ has
a well defined skyrmion number 4.
The $T_d$ isomer of $C_{84}$ has the symmetry of a tetrahedron.
In contrast to the above systems, the
ground state spin configuration for $C_{84}$ has a lower symmetry
than that of the molecule.  The ground state energy is $-113.892689 J$,
which does not reach the lower bound $E_b^{(84)}=-114.541020 J$.
The ground state has skyrmion number 1, which interestingly
is the same skyrmion number as the only other system investigated
with $T_d$ symmetry, the truncated tetrahedron $C_{12}$.

We do not yet understand the $C_n$ problem well enough
to predict the result for a general $n$ isomer without
doing the full calculation.  The results on $C_{20}$, $C_{60}$,
$C_{70}$, and $C_{84}$ are consistent with the hypotheses that
(1) $C_{60}$ is the unique fullerene
that reaches the energy lower bound, and (2) the ground
state has a nonzero skyrmion number when it can be defined.


We now calculate the response of $C_{60}$ to
a magnetic field
by adding
a term $-h \hat z \cdot \sum _ i \vec \sigma _ i $
to the Hamiltonian.
There is a first-order metamagnetic transition, which is
surprising for an isotropic Heisenberg model.
Usually magnetic anisotropy, arising from the spin-orbit
interaction, is required for a metamagnetic transition.
%
The symmetry of the ground state is different above and below
the transition.
Below $h_c$, sites $i$ and $j$ that are diametrically opposite each other
have identical spins, $\vec \sigma _i = \vec \sigma _j$.
Above $h_c$ this symmetry is absent, but
sites $i$ and $j$ that are mapped into each other by a
rotation of $\pi$ about the axis through the midpoint of one
non-pentagon bond
have spins
related by $( \sigma _i^x, \sigma _i^y, \sigma _i^z) =
(- \sigma _j^x, - \sigma _j^y, \sigma _j^z)$.

In nonzero field, the spins in a given pentagon are not coplanar.
The pentagon spin plane normal $\hat n$ is generalized so that
a variable $\hat n_{i,j}$ occupies each pentagon bond,
with $\hat n_{i,j} \sim \vec \sigma _ i \times \vec \sigma _ j $.
The skyrmion number can be calculated for the field
$\hat n_{i,j}$, and it is found that the skyrmion number is 7 both above
and below the transition.


The above results are for the lowest energy state at
a given magnetic field.  One can also obtain
hysteresis loops by following metastable states
while slowly changing the magnetic field,
Fig.\ (\ref{hysteresis}).
A hysteresis loop can be complicated
because of the large number of local minima, but it
never encloses the origin.

We have also investigated the response of $C_{12}$,
$C_{20}$, $C_{70}$, and $C_{84}$
to an external magnetic field.  Of these, only $C_{20}$
has a metamagnetic transition at which the
magnetic moment is discontinuous.  The other members
of this group have a transition at which the moment $M$
is continuous, but $dM/dh$ is discontinuous.
We do not understand this difference in behavior.

The above $M(h)$ calculations for $S = \infty$ do not directly apply to
$C_{60}$, which has $S = 1/2$.  Since $S_z$ is a good
quantum number for the $S = 1/2$ system, it cannot
have a magnetization that is linear in $h$ for small $h$
as shown in Fig. (2).  The closest it can come to Fig. (2)
is to follow
in a staircase fashion, with a series of first order
transitions at which $S$ increases from 0 to 1 to 2, etc.,
with a larger $\Delta S$ jump at the metamagnetic transition.
Since the transition from $S=0$ to 1 is at unobservably
large fields (order of $J$ or 1000 T), the spin susceptibility
vanishes.
For $U >> t$, the orbital susceptibility falls as
$t^5 / U^4$ from the five-membered rings.
Since the physical $U$ is not very much larger than $t$, there is
an orbital contribution to the measured susceptibility.
Elser and Haddon have estimated that the orbital contribution
nearly vanishes for $C_{60}$ with $U=0$.\cite{elser}
The effect of nonzero $U$ is hard to estimate for $C_{60}$,
but we note that for an isolated six-membered ring the orbital susceptibility
for $U/t=4$ is 0.49 that for $U=0$.  The orbital
and spin susceptibility should thus be very small for
$C_{60}$, which is consistent with measurements.\cite{haddon}


We now discuss the $S = 1/2$ wavefunction.
For $S = 1/2$,
$ \sigma$ in Eq. (\ref{ham}) is a $2 \times 2$ matrix.
The simplest prescription
to make an $S = 1/2$ trial wavefunction
is to
form a coherent state
$|\psi _ 0>$ that is a product of spinors.
The spinor on each site $j$ is
quantized in the local $+ \hat z _ j$ direction given by the classical
spin direction.  The expectation $< \psi _ 0 | H | \psi _ 0 >$ is equal
to the classical energy.  Two modifications to this prescription
are required.  The first
is to add zero point spin fluctuations, which lower the energy
and result in a wavefunction $|\psi _ 1>$.
The second is to make the wavefunction the sum over coherent
states representing all of the classical ground states (including
global rotations and parity).  This results in a trial state
of total spin $S=0$.

The spin fluctuation energy is estimated in leading (second)
order perturbation theory.  This calculation gives an
extremely accurate energy for the
square lattice antiferromagnet.\cite{manousakis}
The energy shift for $C_{60}$ is
\begin{equation}
\Delta E = \sum _ j  {{|<j | H | \psi _ 0 > | ^ 2}
\over {\epsilon _0 - \epsilon _j}},
\label{pt}
\end{equation}
where each $|j>$
has two adjacent spins flipped with respect to $|\psi _ 0>$.
For $J_1=J_2=1$, the ground state energy is shifted
to $E_g = E_0 + \Delta E $ $ = -78.541 - 45.676 $ $ = -124.217.$
The correction to the classical energy is
somewhat larger than that for the S=1/2 Heisenberg model on a
square lattice.
The reason is that in this case each site is only three-fold
coordinated and the neighboring spins are not all exactly antialigned,
so that the energy denominators are smaller.
Another attractive variational trial state $| \psi _ 2 >$ is the
product of singlets on
each non-pentagon bond.\cite{rice}  This state has
an energy $E_2= -90$, which is considerably higher than $E_g$.
(The energy $E_2$ can also be reduced by adding fluctuations perturbatively.)
We also calculated the moment reduction in second order
perturbation theory on the coherent state.
The local moment is reduced from 1 to .5590.
This moment is smaller
than that obtained for the square lattice.\cite{manousakis}


Due to quantum fluctuations, a spin 1/2 wavefunction does
not have a unique skyrmion number.
The wavefunctions $|\psi _ 0 >$ and $|\psi _ 1 >$ that we propose
do, however, have unusual, nonvanishing
spin-spin correlation functions even for widely
separated spins on the molecule.  Some of the correlation
functions result from the fact that each pentagon
tends to have a unique normal, which may be calculated
from any of the five adjacent spin pairs in the pentagon,
$\vec \sigma _ i \times \vec \sigma _ j$.  Other nonzero correlation
functions arise because the normals to {\it different}
pentagons are related.
We have calculated some correlation functions
from exact
diagonalizations of the smaller model system, the
truncated tetrahedron.
These calculations indicate that both short and long range classical
spin correlations survive quantum fluctuations.\cite{coffey}
In contrast, the proposed local singlet
state $| \psi _ 2 >$ and the RVB state\cite{kivelson}
have only short range
antiferromagnetic correlations.

Motivated in part by the cuprate superconductors,
there has been a large
effort in calculating the properties of extra holes or electrons
in a planar antiferromagnet using $t-J$ and $t-t'-J$ models.\cite{sat}
An electron doped crystal of $C_{60}$ molecules
is also a superconductor,\cite{hebard} and it
may be useful to do similar calculations in its more
complicated spin field.


In conclusion, we have investigated the low-energy magnetic
properties of undoped $C_{60}$,
the related fullerenes $C_{20}$, $C_{70}$, and $C_{84}$, and
the truncated tetrahedron $C_{12}$
in the strong interaction limit.
We argue that there is no phase transition in $C_{60}$ as a function
of Hubbard $U$, and thus the spin-spin correlation functions at intermediate
$U$ are expected to be similar but smaller than those for large $U$.
In the classical
approximation, the connectivity of the ball leads to an exotic
magnetic ground state with nontrivial topology even for the simplest
antiferromagnetic Heisenberg Hamiltonian.
$C_{60}$ is the only fullerene investigated that
reaches the energy lower bound, meaning that it
has no frustration beyond that
of an elementary pentagon.
The spin correlations in $C_{60}$ may be measurable
by inelastic
neutron scattering, by magnetic x-ray scattering, or by
$^{13}C$ NMR.
%

The authors would like to thank A.~Balatsky, S.~Doniach,
P.~Ginsparg, M.~Inui,
R.~L.~Martin, T.~M.~Rice, D.~Rokhsar,
L.~Sham, R.~Silver, and J.~Thompson for
useful conversations. This work was supported by the US Department of
Energy.


\figure{ (a) The $C_{60}$ molecule is flattened into
a plane by stretching the pentagon on the south pole.
The spin directions are not changed.  The spins in the center
(north) pentagon are in the plane of the figure.  A spin
pointing up out of the plane of the figure is foreshortened,
with an enlarged head.  A spin pointing down into the plane
has an enlarged tail.
Non-pentagon bonds are dotted.
(b) Perspective view of the spin arrangement on the ball.
\label{spin} }

\figure{Typical hysteresis loop, showing magnetic moment $M$ as a function
of applied field $h$, beginning and ending at $h=0$.
The couplings are $J_1=1$ and $J_2=1.2$.
\label{hysteresis} }


\begin{references}
\bibitem[4]{largeu} The evidence for the large $U$ quantum numbers
of the $S= 1/2$ problem is that (1) we have exactly solved the
qualitatively similar 12 site truncated tetrahedron, which
has $L=S=0$ in the large $U$ limit, and (2) the
classical ($S=\infty$) ground state for $C_{60}$ has $\vec S = 0$.
\bibitem[6]{skyrmion} The skyrmion number cannot be defined
directly for the spin field $\vec \sigma$, because
many neighboring spins are antiparallel, which results in
an undefined solid angle.
\end{references}
\end{document}